\documentclass{elsart}

\usepackage[square,comma]{natbib}
\usepackage{graphicx}
\usepackage{pxfonts}
\usepackage{lineno}

\usepackage{amssymb}
\usepackage{footnote}
\usepackage{multirow}
\usepackage{varwidth}
\usepackage{array}
\usepackage{epstopdf}
\usepackage{url}

\journal{}

\begin{document}
\thispagestyle{empty}
\begin{Large}
\textbf{DEUTSCHES ELEKTRONEN-SYNCHROTRON}

\textbf{\large{Ein Forschungszentrum der
Helmholtz-Gemeinschaft}\\}
\end{Large}

DESY 16-079

May 2016

\begin{eqnarray}
\nonumber &&\cr \nonumber && \cr \nonumber &&\cr
\end{eqnarray}
\begin{eqnarray}
\nonumber
\end{eqnarray}
\begin{center}
\begin{Large}
\textbf{A Critical Experimental Test of Synchrotron Radiation Theory with 3rd Generation Light Source}
\end{Large}
\begin{eqnarray}
\nonumber &&\cr \nonumber && \cr
\end{eqnarray}

\begin{large}
Gianluca Geloni,
\end{large}
\textsl{\\European XFEL GmbH, Hamburg}
\begin{large}

Vitali Kocharyan and Evgeni Saldin
\end{large}
\textsl{\\Deutsches Elektronen-Synchrotron DESY, Hamburg}
\begin{eqnarray}
\nonumber
\end{eqnarray}
\begin{eqnarray}
\nonumber
\end{eqnarray}
ISSN 0418-9833
\begin{eqnarray}
\nonumber
\end{eqnarray}
\begin{large}
\textbf{NOTKESTRASSE 85 - 22607 HAMBURG}
\end{large}
\end{center}
\clearpage
\newpage

\begin{frontmatter}



\title{A Critical Experimental Test of Synchrotron Radiation Theory with 3rd Generation Light Source}


\author[XFEL]{Gianluca Geloni,}
\author[DESY]{Vitali Kocharyan,}
\author[DESY]{Evgeni Saldin}
\address[XFEL]{European XFEL GmbH, Hamburg, Germany}
\address[DESY]{Deutsches Elektronen-Synchrotron (DESY), Hamburg, Germany}

\begin{abstract}
A recent "beam splitting" experiment at LCLS apparently demonstrated that after a microbunched electron beam is kicked on a large angle compared to the divergence of the FEL radiation, the microbunching wave front is readjusted along the new direction of motion of the kicked beam.  Therefore, coherent radiation from an undulator placed after the kicker is emitted along the kicked direction without suppression. This strong emission of coherent undulator radiation in the kicked direction cannot be explained in the framework of conventional synchrotron radiation theory. In a previous paper we explained this puzzle. We demonstrated that, in accelerator physics, the coupling of fields and particles  is based, on the one hand, on the use of results from particle dynamics treated according to the absolute time convention and, on the other hand, on the use of Maxwell equations  treated  according to the standard (Einstein) synchronization convention. Here lies the misconception which led to the strong qualitative disagreement between theory and experiment. After the "beam splitting" experiment at LCLS,  it became clear that the conventional theory of synchrotron radiation cannot ensure the correct description  of coherent and spontaneous emission from a kicked electron beam, nor the emission from a beam with finite angular divergence, in an undulator or a bending magnet. However, this result requires further experimental confirmation. In this publication we propose an uncomplicated and inexpensive experiment to test synchrotron radiation theory at 3rd generation light sources.
\end{abstract}

%
%

%
\end{frontmatter}



\section{ Introduction }

The results of the "beam splitting" experiment at LCLS \cite{NUHN}, demonstrated that even the direction of emission of coherent undulator radiation is beyond the predictive power of conventional synchrotron radiation (SR) theory. That experiment constituted the original basis for arguing that there must be something wrong with the theory. In fact, in open contrast with it, after a microbunched electron beam is kicked on a large angle compared to the divergence of the FEL radiation, powerful emission of coherent radiation has been observed in the kicked direction \cite{NUHN}. In a previous paper \cite{OURS} we demonstrated that the effect of aberration of light supplies the basis for a quantitative description of this phenomenon. Maxwell's theory is usually treated under the standard time order, which is based on the use of clocks in uniform motion with respect to the laboratory frame (i.e. in rest with respect to the uniform motion imprinted by the kick) and synchronized by light-signals. In contrast to this, particle dynamics is usually based on a different time order (non-standard, or absolute-time synchronization convention), which is based on the use of clocks at rest with respect to the laboratory frame and synchronized by light-signals. This essential point has never received attention in the physics community. There are two possible ways of coupling fields and particles in this situation. The first, Lorentz's  prerelativistic way, consists in a "translation" of Maxwell's electrodynamics to the absolute time world-picture. The second, Einstein's way,  consists in a "translation" of particle tracking results to an electromagnetic world-picture (that is, to the standard time order). Conventional particle dynamics shows that the electron beam direction changes after the kick, while the orientation of the microbunching phase front stays unvaried.  We showed \cite{OURS1} that a "translation" of particle tracking results to the electromagnetic world picture (or vice versa) predicts a surprising effect, in complete contrast to the conventional treatment. Namely, in the ultrarelativistic asymptotic ($v \to c$), the orientation of the plane of simultaneity (that is the microbunching phase front)  is always perpendicular to  the beam velocity. This effect explains the production of coherent undulator radiation from a modulated electron beam in the kicked direction without the suppression usually predicted by theory.

Immediately after the publication of the results of the "beam splitting" experiment at the LCLS,  it became clear to us that the conventional SR theory cannot even ensure a correct description  of spontaneous emission (i.e. of single particle radiation). In particular, one of the immediate consequences of the beam splitting experiment at LCLS is the occurrence of a red shift of the resonance wavelength, which arises in the kicked direction. Clearly, the conventional SR theory predicts a zero red shift for a fundamental reason related to the Doppler effect. In fact, before and after the kick, the electron has the same speed.  Since in terms of conventional SR theory an electron emits spontaneous undulator radiation without red shift in the kicked direction, we insist on performing an uncomplicated and inexpensive experiment at 3rd generation light sources that can confute the conventional theoretical approach.

In this work we stress that the presence of red shift in undulator radiation automatically implies the same problem for conventional cyclotron radiation theory. In fact, the conventional theory predicts that there should be no red shift for radiation emitted by an electron with velocity directed along and across the magnetic lines of force. In the ultrarelativistic limit well-known analytical formulas which describe the spectral and angular distribution of cyclotron radiation emitted by an electron moving in a constant magnetic field, having a non-relativistic velocity component parallel to the field, and an ultrarelativistic velocity component perpendicular to it. According to the conventional approach, exactly as for the undulator case, the angular-spectral distribution of radiation is a function of the total velocity of the particle due, again, to the Doppler effect. At present, relativistic cyclotron radiation results are textbook examples (e.g. \cite{EC, G, PE} )  and do not require a detailed description. We note, however, that cyclotron-synchrotron radiation emission is one of the most important processes in plasma physics and astrophysics and the results of an experimental test of conventional SR theory  would constitute a truly critical experimental test for a much wider part of  physics than that of SR or XFEL sources.

\section{A crucial experiment to test SR theory with 3rd generation light source}

As already discussed in the Introduction, in this paper we focus on the description of an experiment which can reveal the difference between the predictions of conventional SR theory and our proposed correction in some commonly used setups.

In the following we will assume a filament electron beam approximation, that allows us to take advantage of analytical presentations for single particle SR fields. This means that the emittance of the electron beam is small enough to neglect finite electron beam size and angular divergence in an undulator. In other words,  spatially coherent undulator radiation is assumed to  be used to test SR theory. In accordance with the diffraction condition, the electron beam emittance sets the minimum wavelength at which radiation can be diffraction limited. The performance of synchrotron radiation sources has increased tremendously in terms of spatial coherence during the last years. In particular, new 3rd generation light sources with ultra-low  emittance in the range $\epsilon \simeq 0.3 \mathrm{nm}$, which are presently under operation or under construction, can produce practically fully diffraction limited (i.e. spatial coherent) undulator radiation in the soft X-ray photon energy range around water window (0.25 keV).

We also assume that monochromatization is good enough to neglect the finite bandwidth of the radiation around the fundamental frequency, and that the electron beam energy spread is negligible. Contrarily to the bandwidth of a frequency filter, the energy spread is fixed for a given facility: its  presence constitutes a fundamental effect. The typical energy spread for 3rd generation light sources is of order $0.1 \%$. For the LCLS source this figure is about an order of magnitude smaller. We studied the impact of a finite energy spread using the expression for the intensity of a diffraction limited beam including energy spread. We compared, at the resonance frequency, i.e. at zero detuning parameter, the case for negligible energy spread  with the case corresponding to an energy spread of $0.1 \%$  for a  number of undulator periods $N_w = 70$, which are typical for 3rd generation light sources. We found that differences in the maximum intensities are within $10 \%$. This reasoning allowed us to conclude that the simplest analytical result for zero energy spread can be applied to practical cases of interest involving 3rd generation light sources and undulators with up to $70$ periods with good accuracy.

When one needs to specify the angular-spectral flux density at any position down the beam line, one needs to calculate the field at any position down the beam line. In order to do so, we first discuss the field from a single relativistic electron moving without kick along the undulator axis. We indicate the velocity of the electron with $v$ and the Lorentz factor with $\gamma = 1/\sqrt{1-v^2/c^2}$. We consider a planar undulator, so that the transverse velocity of the electron can be written as

\begin{equation}
	\vec{v}_\bot(z) = - {c K\over{\gamma}} \sin{\left(k_w z\right)}
	\vec{e}_x~, \label{vuzo}
\end{equation}
where $k_w = 2\pi/\lambda_w$, with $\lambda_w$ the undulator period and $K$ the undulator parameter

\begin{equation}
	K=\frac{\lambda_w e H_w}{2 \pi m_\mathrm{e} c^2}~, \label{Kpara}
\end{equation}
$m_\mathrm{e}$ being the electron mass and $H_w$ being the maximum of the magnetic field produced by the undulator on the $z$ axis. The resonance condition at the fundamental harmonic is given by

\begin{eqnarray}
\frac{\omega}{v_z} - \frac{\omega}{c} =
k_w ~, \label{rsfirsth1}
\end{eqnarray}
where $v_z$ is the average speed in the $z$ direction. In terms of $K$ and $\gamma$,  the resonance frequency is

\begin{eqnarray}
\omega_r =
\frac{2c k_w\gamma^2}{1+K^2/2 }~. \label{rsfirsth}
\end{eqnarray}
SR theory is naturally developed in the space-frequency domain, as one is usually interested in radiation properties at a given position in space at a certain frequency. Here we define the relation between temporal and frequency domain via the following definition of Fourier transform pairs:

\begin{eqnarray}
&&\bar{f}(\omega) = \int_{-\infty}^{\infty} dt~ f(t) \exp(i \omega t ) \leftrightarrow
f(t) = \frac{1}{2\pi}\int_{-\infty}^{\infty} d\omega \bar{f}(\omega) \exp(-i \omega t) ~.
\label{ftdef2}
\end{eqnarray}
A well-known expression for the angular distribution of the first harmonic field in the far-zone can be found in literature. Such expression is axis-symmetric, and can therefore be presented as a function of a single observation angle $\theta$, where

\begin{equation}
	\theta^2 = \theta_x^2+\theta_y^2~, \label{thsq}
\end{equation}
$\theta_x \simeq \tan \theta_x = x_0/z_0$ and $\theta_y \simeq \tan \theta_y = y_0/z_0$ being angles measured from the undulator $z$-axis in the horizontal and in the vertical direction, and $(x_0,y_0,z_0)$ being the observation position relative to the center of the undulator. In the following we will always assume that the ultra-relativistic approximation is satisfied, which is the case for SR setups. As a consequence, the paraxial approximation applies too. The paraxial approximation implies a slowly varying envelope of the field with respect to the wavelength. It is therefore convenient to introduce the slowly varying envelope of the transverse field components as

\begin{equation}
\widetilde{E}(z,x,y,\omega) = \bar{E}(z,x,y,\omega) \exp{\left(-i\omega z/c\right)}~. \label{vtilde}
\end{equation}

One then obtains the following distribution for the slowly varying envelope of the electric field:

\begin{eqnarray}
	&&{\vec{\widetilde{E}}}(\theta) = -\frac{K \omega e
		L  }{2 c^2 z_0 \gamma} A_{JJ}\exp\left[i\frac{\omega \theta^2
		z_0}{2c}\right] \mathrm{sinc}\left[\frac{L}{2}\left(C+\frac{\omega
		\theta^2}{2c} \right)\right] \vec{e}_x~,\cr && \label{generalfin4}
\end{eqnarray}
where $\mathrm{sinc}(\cdot) \equiv \sin(\cdot)/(\cdot)$. The field is polarized in the horizontal direction.  Here $L = \lambda_w N_w$ is the undulator length and  $N_w$ the number of undulator periods, $\omega = \omega_r + \Delta \omega$, $C =  k_w \Delta\omega/\omega_r$ and $\omega_r$ is the fundamental resonance frequency  which can be obtained from Eq. (\ref{rsfirsth}). Finally, $A_{JJ}$ is defined as

\begin{equation}
	A_{JJ} = J_o\left(\frac{K^2}{4+2K^2}\right)
	-J_1\left(\frac{K^2}{4+2K^2}\right)~, \label{AJJdef}
\end{equation}
$J_n$ being the n-th order Bessel function of the first kind. Eq. (\ref{generalfin4}) is valid only  when the resonance approximation is valid. This approximation does not replace the paraxial one, based on $\gamma^2 \gg 1$, but it is used together with it. It takes advantage of another parameter that is usually large, i.e. the number of undulator periods $N_w \gg 1$.

The physical meaning of Eq. (\ref{generalfin4}) can be best understood by considering the properties of radiation in terms of photon flux density. For a filament electron beam of current $I$, the angular spectral flux density in the direction $(\theta_x, \theta_y)$  can be written as

\begin{eqnarray}
	\frac{d F}{d \Omega} = \frac{d \dot{N}_{ph}}{d \Omega (d \omega/\omega)}  = \frac{I}{e \hbar} \frac{c z_0^2}{4\pi^2} |\widetilde{E}|^2 ~,
	\label{dfdom}
\end{eqnarray}
where $d\dot{N}_{ph}/(d \omega/\omega)d \Omega$ is the number of photons per unit time per unit solid angle per relative frequency bandwidth, and $\widetilde{E}$ is the slowly varying envelope of the electric field produced by a single electron  in the space-frequency domain \footnote{Note that Eq. (\ref{dfdom}) depends on the units chosen, in this case Gaussian units, and on definition of Fourier transformation, Eq. (\ref{ftdef2})}.

The maximum value of $|\widetilde{E}|^2$ at the resonance wavelength as a function of $\theta_x$ and $\theta_y$ is reached on-axis, i.e. for $\theta_x = \theta_y = 0$. The on-axis angular spectral flux at perfect resonance is given by

\begin{eqnarray}
	\max\left(\frac{dF}{d \Omega}\right) = \frac{I}{e}\alpha K^2 A_{JJ}^2 \frac{L^2}{4\lambda^2\gamma^2}~,
	\label{maxdfdom}
\end{eqnarray}
where $\alpha = e^2/(\hbar c)$ is the fine structure constant. The angle-integrated spectral flux $F = d \dot{N}_{ph}/(d \omega/\omega)$ is defined as

\begin{eqnarray}
	F = \int \frac{dF}{d \Omega} d \theta_x d \theta_y ~.
	\label{angint}
\end{eqnarray}
If we substitute Eq. (\ref{dfdom}) in Eq. (\ref{angint}) we obtain

\begin{eqnarray}
	F = \frac{I}{e} \pi \alpha K^2 A_{JJ}^2 \frac{N_w}{2(1+K^2/2)} ~.
	\label{Fsub}
\end{eqnarray}

Let us now discuss the effect of a deflection angle $\eta \simeq \tan \eta = v_x/v_z \simeq v_x/c$, that is our kick,  in accordance to conventional SR theory. The undulator radiation field emitted by an electron with a  deflection angle can be derived after rather cumbersome calculations. These calculations are based on the well-known formulas for the retarded fields, and were first presented in \cite{KIM1,KIM3}. using our previous notation, an explicit expression for the radiation field emitted by an electron with a detuning from resonance $C$ and a deflection angle $\eta$ is

\begin{eqnarray}
	&&{\vec{\widetilde{E}}}( \vec{\theta}, \vec{\eta}) = -\frac{K \omega e
		L  }{2 c^2 z_0 \gamma} A_{JJ}\exp\left[i\frac{\omega \theta^2
		z_0}{2c}\right] \mathrm{sinc}\left[\frac{L}{2}\left(C+\frac{\omega
		\left| \theta - \eta\right|^2}{2c} \right)\right] \vec{e}_x~. \label{generalfin5}
\end{eqnarray}

It is possible to give an elementary explanation of one obvious effect due to electron deflection. Since the magnetic field experienced by the electron is assumed to be independent of its transverse coordinate, the trajectory followed is still sinusoidal, but the effective undulator period is now given by $\lambda_w/\cos \eta \simeq (1+\eta^2/2) \lambda_w$. This induces a relative red shift in the resonant wavelength $\Delta \lambda/\lambda = \eta^2/2$. In practical cases of interest we may estimate a deflection angle up to about $\eta \simeq 1/\gamma$. Then, the relative red shift $\Delta \lambda/\lambda \simeq 1/\gamma^2$ should be compared with the relative bandwidth of the resonance, that is $\Delta \lambda/\lambda \sim 1/N_w$, $N_w$ being the number of undulator periods. In all situation of practical relevance $1/\gamma^2 \ll 1/N_w$ and, as a result, the red shift in the resonance wavelength due to the increase of the effective undulator period can be neglected.

It is clear from the above that, according to conventional SR theory, if we consider radiation from one electron  (or from a filament beam) at detuning $C$ from resonance, the introduction of a kick  only amounts to a rigid rotation of the angular distribution along the new direction of the electron motion. This is plausible, if one keeps in mind that  after the kick the particle has the same energy and emits radiation in the kicked direction  owing to the Doppler effect. After such rotation, Eq. (\ref{generalfin4}) transforms into Eq. (\ref{generalfin5}).

Let us now discuss the effect of a kick in accordance to our predictions \cite{OURS}. Lorentz demonstrated  that the principle of Galilean relativity, i.e. the Galilean law of addition of velocities, holds not only in mechanics but also in  experiments involving optics and electrodynamics as long as we neglect effects proportional to $v_x^2/c^2$. This accuracy is more than sufficient in our case of  interest, because the kick angle ($\eta \lesssim 1/\gamma$)  is very small (order of 0.1 mrad), so that a first approximation over the parameter $v_x/c$  yields a correct quantitative description. In what follows we will thus endorse a Galilean description. According to the principle of Galilean  relativity, the physical laws  appear the same in any two inertial frames, $S_0$ and $S$, and are related by a Galileo transformation. In the inertial frame $S_0$, that is the laboratory frame, the electron velocity components after the kick are $(v_x, 0, v_z)$,  where $v_z = \sqrt{v^2 - v_x^2}$ and $v$ is the beam velocity along the $z$-axis, i.e. the undulator axis upstream of the kicker. Consider now an inertial reference frame $S$ moving with uniform motion at speed $v_x$ along the $x$-axis of the frame $S_0$. In  $S$, the electron velocity components are just $(0,0,v_z)$. Since all inertial frames are equivalent, $S$ and $S_0$  can only be distinguished from one another by their relative speed $v_x$, and we can transform quantities from $S$ to $S_0$ and viceversa using a Galileo transformation. Now we remind that since we resort to Galilean relativity, simultaneity of events does not depend on the reference frame chosen ($S$, $S_0$ or any other inertial frame), that is it is absolute. Therefore, the kicked electron, as seen in the frame $S$, reproduces exactly the situation upstream of the kicker as seen in the frame $S_0$. However, it is clear from the above that if an electron is at perfect resonance in $S_0$ before the kick and moves with longitudinal velocity $(0,0,v)$, then after the kick  the same  electron  cannot be  at perfect resonance in the frame $S$, because of  the different longitudinal velocity $v_z$. We find that the red shift in the resonance wavelength due to the difference between longitudinal velocities  cannot be neglected in practical situation. The rule for computing the result in laboratory frame $S_0$ is simple. One takes the velocity of the light with respect to the frame $S$ and adds it vectorially to the velocity of the frame $S$ with respect to the laboratory frame $S_0$. The direction of the resulting vector is the apparent direction of the source of light as measured at the observer position in the laboratory frame. The rule conforms to the principle of Galilean kinematics\footnote{We thus describe a complicated situation by finding a reference system where analysis is easier and then we transform back to the old reference frame. Note that leading treaties and textbooks on classical electrodynamics tell us that a Lorentz transformation always leads to correct result. However, if we apply a Lorentz boost, our predictions of radiation properties  will be incorrect in our case: it seems puzzling that only a Galileo boost is correct. However, we note that even in the non-relativistic limit, when we can neglect second order  corrections in $v_x/c$, which are intrinsically relativistic, Lorentz and Galilean transformations are different. In fact the term $x v_x/c^2$ in the Lorentz transformation for time, leading to relativity of simultaneity,  is a first order correction. Yet this term is only conventional and has no direct physical meaning: with a suitable clocks synchronization convention this term can be eliminated. In other words, Galilean  and Lorentz transformations are different even in the non-relativistic limit only because they are based on the use of different synchronization conventions. Our point is that the usual description of electron motion is based on the use of clocks in rest relative to the laboratory frame, synchronized by light-signals,  and that under this convention only a Galilean boost can be used. In contrast, Lorentz transformation is based on the use clocks in rest relative to the frame moving with translational velocity $v_x$, synchronized by light-signals.  More details can be found in section \ref{disc}.}.

The preceding Galilean approach gives the following expression for the radiation field in question

\begin{eqnarray}
&&{\vec{\widetilde{E}}}( \theta, \eta) = -\frac{K \omega e
	L  }{2 c^2 z_0 \gamma} A_{JJ}\exp\left[i\frac{\omega \theta^2
	z_0}{2c}\right] \mathrm{sinc}\left[\frac{L}{2}\left(C+ \frac{\omega\eta^2}{2c} + \frac{\omega
	\left| \theta - \eta\right|^2}{2c} \right)\right] \vec{e}_x~.\cr && \label{generalfin6}
\end{eqnarray}

\begin{figure}
\begin{center}
\includegraphics[width=0.6\textwidth]{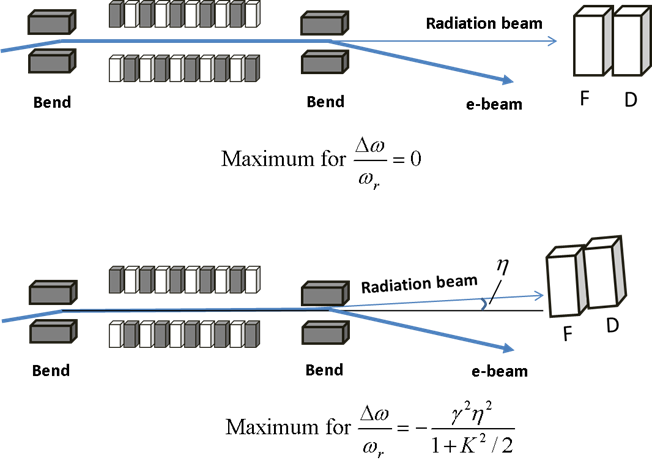}
\end{center}
\caption{Basic setup for our proposed critical test. Top: the case for a filament beam without kick. Bottom: the case for a filament beam kicked by an angle $\eta$. In both cases, the X-ray pulse is filtered by a monochromator $F$ and the total energy recorded by a detector $D$ as a function of the detuning.} \label{F3}
\end{figure}
This formula has nearly, but not quite the same form as Eq. (\ref{generalfin5}), the difference consisting in the term $\omega\eta^2/(2c)$ in the argument of the $\mathrm{sinc}$ function. Attention must be called to the difference in resonance frequency between undulator radiation setups with and without the kick. When the electron has a uniform translational motion with  velocity $v_x$, the red shift in the resonance frequency can be expressed by the formula $\Delta\omega/ \omega_r =  - \omega_r\eta^2/(2k_w c)$, which can be written as $\Delta\omega/ \omega_r =  - \gamma^2\eta^2/(1+K^2/2)$. We now see a second order correction $\eta^2$ that is, however, multiplied by a large factor $\gamma^2$. We thus conclude that both the apparent readjusting of microbunch front orientation and red shift in the resonance wavelength observed at the LCLS beam splitting experiment can be quantitatively explained with the help of Galilean relativity.

In order to confirm our prediction, we propose a simple experiment at 3rd generation light sources with ultra-low emittance in the soft X-ray range. The basic setup for a test experiment is sketched in Fig. \ref{F3}. The soft X-ray undulator beam line should be tuned to a minimum photon energy (typically this limit is related with the water window). The radiation pulse goes through a monochromator filter $F$  and its energy is subsequently measured by the detector $D$. No precise monochromatization of the undulator radiation is required in this case: a monochromator line width $\Delta \omega/\omega \simeq 0.001$ is sufficient. In order for our test experiment to be carried out, it is necessary to control the beam kicking e.g. by corrector magnets.  In the case of no kick the maximum pulse energy registered by the detector will coincide with the monochromator line  tuned to resonance, Fig. \ref{F3} (top). When the kick is introduced, Fig. \ref{F3} (bottom), the conventional SR theory still predicts (up to corrections $\Delta \omega/\omega_r = \eta^2/2$, discussed before and negligible) a zero red shift in the resonance wavelength. In contrast to this, one of the immediate consequences of our theory is the occurrence of a non-zero red shift of the resonance wavelength, which arises because the electron beam has a uniform translational motion with velocity $v_x \simeq v\eta$ in the direction perpendicular to the undulator axis. The object of the whole experiment is to confirm or confute the difference in resonance frequencies between the two configurations in Fig. \ref{F3}. The proposed experimental procedure is reasonably simple and based on  relative measurements. A necessary condition for this type of test experiment is the validity of the filament beam approximation.

\section{\label{disc} Discussion and Conclusions}

\begin{figure}
\begin{center}
\includegraphics[width=0.8\textwidth]{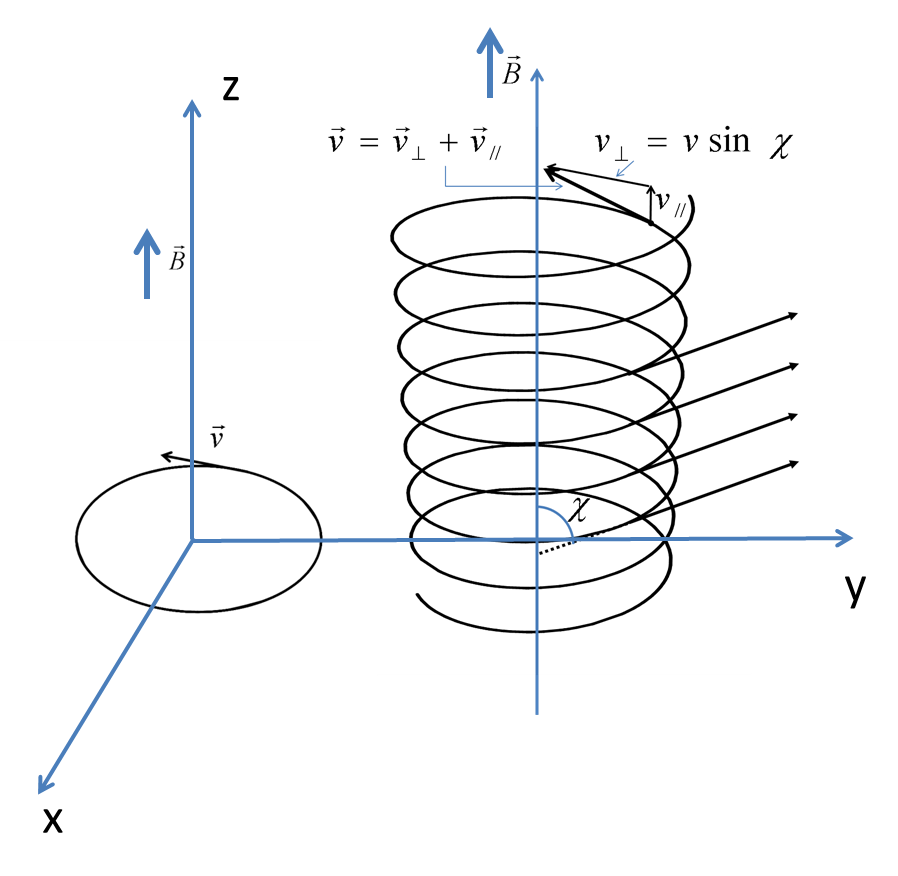}
\end{center}
\caption{Geometry for radiation production from circular and helical motion. On the same reference system we show an electron in a uniform magnetic field (left)  on a circular orbit and (right) on a helical orbit.} \label{F4}
\end{figure}

There is another interesting problem where our correction of SR theory results is required. It is the problem of cyclotron radiation. Calculations pertaining radiation from an ultra-relativistic electron in helical motion in a uniform magnetic field are well-known, see e.g. \cite{EC, G, PE}. Similarly as for the case of undulator radiation, however, these calculations lead to incorrect results because of improper analysis for the case of a uniform translational motion of the electron in the direction of magnetic field. Consider Fig. \ref{F4} (right). We argue that the  angular divergence and critical frequency of the radiation emitted by an electron with relativistic factor $\gamma = 1/\sqrt{1-v^2/c^2} \gg 1$ spiraling with pitch angle angle $\chi \simeq \pi/2$  about a magnetic field $B$ are not $\Delta \theta \sim 1/\gamma$ and $\omega_c \sim B\gamma^2$ as usually stated but rather, respectively, $\Delta \theta \sim 1/\gamma_\perp$  and $\omega_c \sim B\gamma^2_\perp$, where $\gamma_\perp = 1/\sqrt{1-v_\perp^2/c^2}$ and $v_\perp = v\sin \chi$.

The influence of the uniform translational motion along the magnetic field direction on the radiation emission can be described purely kinematically. Up to date, our community never considered the question of the method of synchronizing spaced clocks used for describing the particle motion. However, without a clear definition of this method, any attempt to solve electrodynamic problems  will lead to incorrect results. In our previous paper \cite{OURS1} we demonstrated that Galilean transformations supply the basis for a quantitative description of the results obtained at LCLS. Since the creation of special relativity, most researches assume that Lorentz transformations immediately follow from the postulates of the theory of relativity. However these postulates alone are not sufficient to obtain Lorentz transformations:  one additionally needs to synchronize  spatially separated moving clocks  with the help of light signals. If this is done using the Einstein synchronization convention, Lorentz transformations follow. However, if the same clocks are synchronized following a different synchronization convention, other transformations are obtained. The main point behind our criticism of conventional SR theory is  that, under a uniform translation, the electron motion is not described by using clocks in rest relative to the moving reference frame and synchronized by light-signals, but rather by using clocks in rest relative to the laboratory frame and synchronized by light-signals, i.e. following the so-called absolute time convention. This statement follows from the observation that particle dynamics in the laboratory frame treats time as an independent variable. Unfortunately, in the work of the majority of researches dealing with cyclotron theory development, the synchronization procedure of clocks has either not been considered at all or has been performed incorrectly.

Widely accepted expressions for the angular and spectral distributions of radiation from an ultra-relativistic electron on a helical orbit were calculated in \cite{WES, EP}. Let us discuss in some detail the cyclotron radiation emitted by an  electron moving in a constant magnetic field with a non-relativistic component of the velocity parallel to the direction of the magnetic field, and a ultra-relativistic component perpendicular to it. Here we shall only give  some final results and discuss their relation with the conventional SR theory from a bending magnet with reference to Fig. \ref{F4}. In the case of a uniform translational motion with non-relativistic velocity along the magnetic field direction, the radiation field  in the far zone according to \cite{G}, and using the notations in that reference, is given by

\begin{eqnarray}
{\vec{\widetilde{E}}}(\chi, \alpha)  &&  \sim
\Bigg\{ \vec{e}_x
\left[(\xi^2 + \psi^2) K_{2/3} \left(\frac{\omega}{2\omega_c}
\left(1+\frac{\psi^2}{\xi^2}\right)^{3/2}\right)\right]\cr   && - i \vec{e}_y \left[(\xi^2 +
\psi^2)^{1/2} \psi K_{1/3}\left(\frac{\omega}{2\omega_c}
\left(1+\frac{\psi^2}{\xi^2}\right)^{3/2}\right)\right] \Bigg\}~,  \label{EF}
\end{eqnarray}
where $K_{1/3}$ and $K_{2/3}$ are the modified Bessel functions, $\xi = 1/\gamma$, $\psi = \chi - \alpha$ ($\chi$ is the angle between $\vec{v}$ and $\vec{B}$ and $\alpha$ that between $\vec{n}$ and $\vec{B}$); the angle $\psi$ is clearly the angular distance between the direction of the electron velocity $\vec{v}$ and the direction of observation $\vec{n}$. Here $\omega_c$ is defined by $3e B\gamma^2/(2m_ec)$.

This result can be understood most easily by comparison with the theory of the radiation emitted by an ultra-relativistic electron moving instantaneously at constant speed on a circular path, Fig. \ref{F4} (left). For convenience we now define the small deflection angle $\eta = \pi/2 - \chi$, and we discuss its effects in accordance with conventional SR theory. It is clear from  Eq. (\ref{EF})  that if we consider the  radiation from an electron with  relativistic factor $\gamma$ moving on a circular orbit (Fig. \ref{F4} (left)),  the introduction of the kick only amounts (as before for the undulator case) to a rigid rotation of the angular distribution along the new direction of the electron motion (Fig. \ref{F4} (right)). As for the undulator case, this is intuitively sound if one keeps in mind that  after the kick the electron has the same velocity in magnitude, and emits radiation in the kicked direction owing to the Doppler effect.

\begin{figure}
\begin{center}
\includegraphics[width=0.8\textwidth]{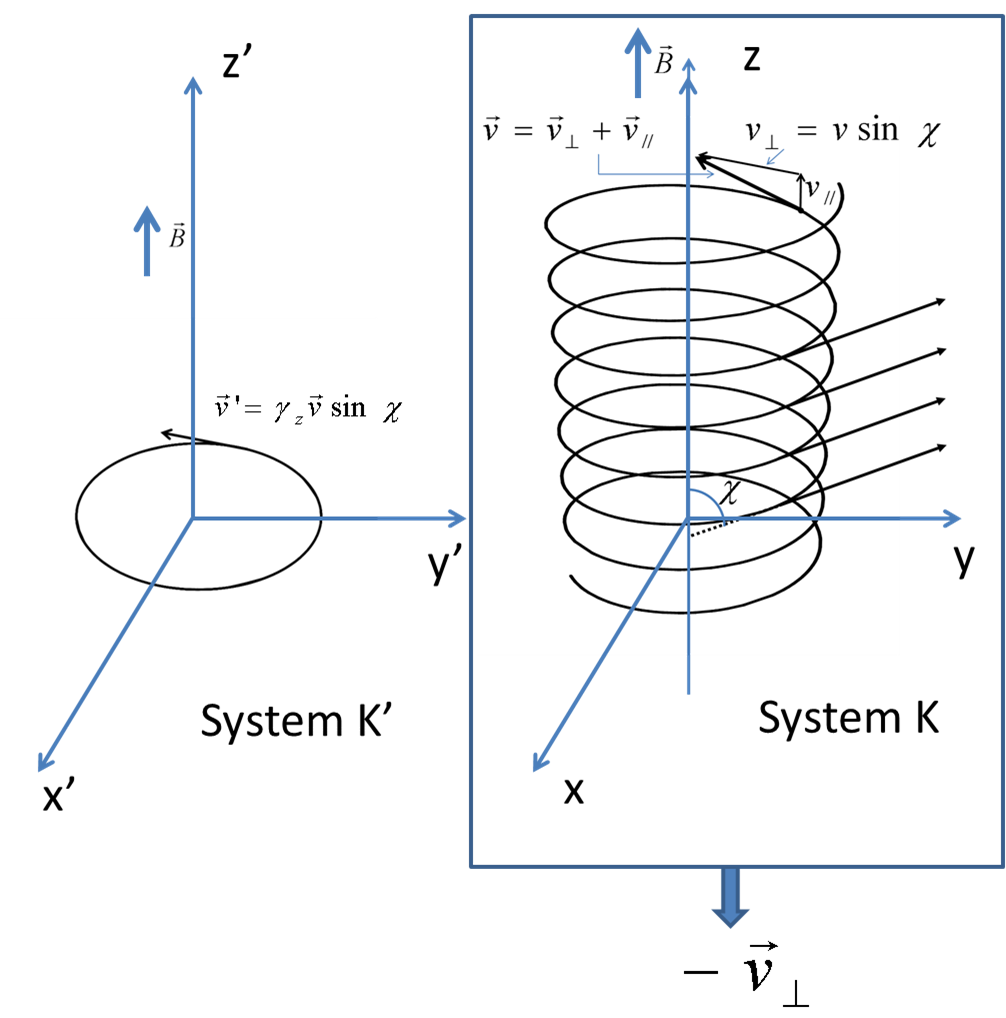}
\end{center}
\caption{Geometry for radiation production from circular and helical motion, following \cite{OS}. In different reference systems we show (left, system $K'$) an electron on a circular orbit and (right, system $K$) the helix of an electron in a uniform magnetic field. System $K'$ is obtained rom system $K$ using a Lorentz boost in the $z$ direction.} \label{F5}
\end{figure}
The angular-spectral distribution in Eq. (\ref{EF}) was recovered in treatments which make no explicit use of the theory of relativity \cite{WES, EP}. When there is motion along the field, that is for $\eta \ne 0$, like in Fig. \ref{F4} (right), the calculation leading to Eq. (\ref{EF}) is rather elaborate.  It is therefore desirable to have an independent derivation explicitly based on the theory of relativity. This was carried out in \cite{OS}. Consider Fig. \ref{F5}. The reference frame $K'$ in which the electron moves in circular motion  was transformed  to a frame $K$, in which the electron proceeds following a helical trajectory  (note that the magnetic field remains unchanged by the transformation from $K'$ to $K$). The reader can compare Fig. \ref{F5} with Fig. \ref{F4}, where both motions happen in the same inertial frame, the laboratory frame. In \cite{OS} it was shown that Eq. (\ref{EF})  holds, indeed, in the frame $K$  for a particle whose velocity is $(v_x,v_y,v_z) = (v\sin \chi\cos \phi, v\sin\chi\sin\phi, v\cos\chi)$. The Lorentz transformation, which leads to the value $v_z = v\cos\chi$ for the $z$-component of the velocity yields

\begin{eqnarray}
(v_x,v_y,v_z) = (v'\cos\phi'/\gamma_z, v'\sin\phi'/\gamma_z, v_z) ~,
\end{eqnarray}
where $\gamma_z =1/\sqrt{1-v_z^2/c^2}$, $v'$ is the velocity of the electron in the frame $K'$ and the phase angle $\phi' = \phi$ is invariant.  This means that, in order to end up in $K$ with a transverse velocity $v_\perp = v\sin\chi$, one must start in $K'$ with $v' = \gamma_z v\sin\chi$. In the ultra-relativistic approximation $\gamma_\perp^2 = 1/(1-v_\perp^2/c^2)\gg 1$, and one finds the simple result $v' = v$, so that a Lorentz boost with non-relativistic velocity $v_z$ leads to a rotation of the particle velocity $\vec{v}$ of an angle $\eta = \pi/2 - \chi \simeq v_z/c $. If one transforms  the radiation field for a particle in circular motion in the system $K'$ and neglects second order terms $v_z^2/c^2 \ll 1$ in observation angle and frequency, one obtains the result that the effect of a kick amounts to a rigid rotation of the angular-spectral distribution of the radiation emitted by an electron moving with velocity $v$ on a circle that is, once more, Eq. (\ref{EF}). This behavior of the radiation field under Lorentz transformations is widely accepted as an independent check of the correctness of Eq. (\ref{EF}).

We now present our reasons why the conventional theory of cyclotron radiation, and in particular Eq. (\ref{EF}), should be modified.

Let us start with the treatment \cite{OS}, which makes explicit use of relativity. When we discuss the case of Fig. \ref{F4} we refer to a single reference frame, and we need to define exactly what we are doing and what we are measuring.  In order to measure the velocity of our electron, we must first synchronize distant clocks in the laboratory frame, which we assume to be inertial. The conventional choice of clock synchronization  corresponds, in our case, to the choice of standard "light-signal" synchronization. After this choice is made, the speed of light measured with these synchronized clocks in the laboratory frame will always be $c$, isotropically, i.e. in any direction of space.  Using these clocks, which are fixed and at rest in the laboratory frame, we measure, in the same laboratory frame, the electron velocity components before (Fig. \ref{F4}, left) and after (Fig. \ref{F4}, right) the kick. In contrast, reference \cite{OS} investigated a Lorentz transformation from the reference frame $K'$ in which the electron moves in circular motion to the moving frame $K$, in which the electron moves in a helical motion, Fig. \ref{F5}.  When we perform this transformation we  must consider that the circular motion and the helical motion are now described in the judgement of two different reference systems, $K'$ and $K$ (Fig. \ref{F5}), with different time coordinates. However, we should remember that what we really need is to discuss circular and helical motion in  the same reference frame (the laboratory frame in Fig. \ref{F4}) where the same clocks at rest are synchronized according to the standard convention for electron velocity  measurements for both setups. If we, at variance, make a Lorentz transformation from $K'$ to $K$ to describe the situation in Fig. \ref{F5}, we automatically change the rhythm of the clocks, and we must take into account that this corresponds to a change in the horizontal velocity component (the horizontal coordinates do not change because the boost is in the vertical direction). This means that the Lorentz transformation described in \cite{OS} has actually no relation with our problem of comparing SR radiation before and after the kick. The changes through the action of a kicker in the dynamical evolution of the electron from Fig. \ref{F4} left to Fig. \ref{F4} right, both in the laboratory frame,  cannot be described as a Lorentz boost from the system $K'$ to the boosted system $K$ in Fig. \ref{F5}.

In the following we present a calculation of radiation from an electron in helical motion based on the solution of Maxwell's equations in the laboratory frame and discuss the influence of a uniform translation along the magnetic field on radiation phenomena. More generally,  we consider a system of electrons  having a common uniform translation, which is characterized by a velocity $\vec{v} = (0,0,v_z)$. We derive the equations that describe this situation  by a change of variables \cite{L}. In fact, it is natural to refer the phenomena in a moving frame that is fixed to the system and shares its translation; these new coordinates will be represented by $x',y',z'$. They are given by $ x' = x, y' = y, z' = z - v_z t$.  For any quantity $F$ (for example the radiation field or the current density), which depends on spatial coordinates and time, we may distinguish two differential coefficients, which we denote respectively as $\partial F/\partial t$ and $(\partial F/\partial t)$. The first partial derivative is used when $F$ is considered as a function of $t$ and the "absolute" coordinates. i.e. when the coordinates with respect to axes fixed in the laboratory frame are taken as independent variables. The second partial derivative is used when $F$ is considered as a function of $t$ and the "relative" coordinates, i.e. when the coordinates with respect to the axes moving with velocity $v_z$ are taken as independent variables. The relation between the two quantities is expressed by the formula

\begin{eqnarray}
	\left(\frac{\partial F}{\partial t}\right) = \frac{\partial F}{\partial t} + v_z \frac{\partial F}{\partial z}~.
\end{eqnarray}
As to the differential coefficients with respect to spatial coordinates we have $\partial/\partial z' = \partial/\partial z$ (and similarly $\partial/\partial x' = \partial/\partial x$ and $\partial/\partial y' = \partial/\partial y$).   Therefore the d'Alambertian, which enters in the basic equation of electromagnetism, is a partial differential operator whose change of form can be calculated just by replacing $\partial/\partial t$ with $(\partial/\partial t) = \partial/\partial t + v_z\partial/\partial z$.

After this replacement, taking $x,y,z$ and $t$ as independent variables, we can see that the inhomogeneous wave equation for the electric field in the laboratory frame has nearly but not quite the usual, standard form that takes when there is no translation. The main difference consists in the "interference" term $2(v_z/c)\partial^2 \vec{E}/\partial t\partial z$.  Once we have the wave equation in the correct form in the laboratory frame, we may apply it to the study of various phenomena. Following \cite{L}, however, the discussion of many cases can be based on a mathematical trick,  without the direct solution of the modified wave equation. Lorentz found that the solution of the electrodynamic problem in the "true" time $t$ can be obtained with minimal efforts by formally desynchronizing the "true" time $t$ to $t' = t-z v_z/c^2$ and using $t'$ without changing the d'Alambertian in the form outlined above. It is immediately seen by direct calculations that a shift of time is necessary in order to the eliminate interference term\footnote{We have just seen that the wave equation in the laboratory frame, i.e. the equation that describes radiation from an electron in helical motion by using $(x,y,z,t)$ as independent variables, includes an "interference" term in ${\partial^2\vec{E}}/({\partial t\partial z}) $, see the left part of Eq. (\ref{inte}), where both derivatives with respect to $t$ and $z$ appear. It follows that, due to the presence of a translational motion, the electron does not radiate like an electron moving on a circle, but it rather picks up extra-effects. In principle, the modified wave equation may be solved directly without mathematical tricks, for example by numerical methods, and one may directly derive wavefront readjusting and red shift associated with the "interference" term.
}:

 \begin{eqnarray}
 && \nabla_{\perp}\vec{E} +	\left(1-\frac{v_z^2}{c^2}\right)\frac{\partial^2\vec{E}}{\partial z^2}  - 2\left(\frac{v_z}{c}\right)\frac{\partial^2\vec{E}}{\partial t\partial z} - \frac{1}{c^2}\frac{\partial^2\vec{E}}{\partial t^2} \cr && =
 \nabla_{\perp}\vec{E} + \left(1-\frac{v_z^2}{c^2}\right)\frac{\partial^2\vec{E}}{\partial z^2}
 - \left(1-\frac{v_z^2}{c^2}\right)\frac{1}{c^2}\frac{\partial^2\vec{E}}{\partial t'}^2~.
 \label{inte}
 \end{eqnarray}
As we continually neglect quantities of second order ($v_z^2 \ll c^2$), we may  substitute $(1-v_z^2/c^2)\partial^2\vec{E}/\partial z^2$ with $ \partial^2\vec{E}/\partial z^2$. The situation is very different in the case of the term  $(1-v_z^2/c^2)\partial^2\vec{E}/\partial (ct')^2$, because this term yields a "renormalized" velocity of light $c' = c/\sqrt{1 - v_z^2/c^2} \simeq c + v_z^2/(2c^2)$, which is now involved, instead of $c$ in the resonance condition with relativistic velocity $v_{\perp}$.

We come to the conclusion that the interference term described above yields an aberration angle $v_z/c$ and an effective increase in the velocity of light from $c$ to $c'$, i.e. a red shift in the resonance condition in accordance with experimental results of the LCLS beam splitting experiment with a modulated electron beam. In order to work with the conventional velocity of light, we can just make consistent use of the substitution $t' \longrightarrow t'/\sqrt{1 - v_z^2/c^2}$. In this case, the electron velocity $v = \sqrt{v_z^2 + v_{\perp}^2}$ will be transformed to $v_{\perp}$. The treatment presented here shows that there are subtle cancelations that actually lead to a simple result. In order to obtain this result, we made use of a famous theorem formulated by Lorentz, the theorem of corresponding states. According to this theorem, an electron (or a system of electrons) in a state of uniform translation with velocity $v_z$  produces radiation with identical angular-spectral density distribution as an electron in the corresponding state without translation. So, the rule for computing the radiation in the case of uniform translation is simple. One takes the radiation characteristics in the corresponding state and introduces an angle of aberration of $v_z/c$ radians, with $v_z \ll c$. In the case of an electron in a helical motion with relativistic factor $\gamma$, the corresponding state is an electron moving in a circle with relativistic factor $\gamma_{\perp}$.  This is plausible if one keeps in mind that radiation is related with acceleration, which is identical in the corresponding states. The influence of the uniform translation is responsible for the aberration of light and is described purely kinematically, as it must be. The theorem of corresponding states demonstrates that the principle of Galilean relativity, i.e. invariance under Galilean transformations, is to  be expected not only in mechanics, but also in experiments involving optics and electrodynamics in the case when intrinsically relativistic effects ( i.e. effects  of order of $v_z^2/c^2$ in radiation properties ) may be neglected. We demonstrated the effectiveness of this broad conclusion above, when the undulator case was discussed.

\end{document}